\documentclass[prb,twocolumn,showpacs,superscriptaddress,floatfix,oneside]{revtex4}%
\usepackage{amssymb}
\usepackage{amsfonts}
\usepackage{graphicx}
\usepackage{epsfig}
\usepackage{amsmath}
\begin{document}
\title{Longitudinal magnetoresistance in Co-doped BaFe$_2$As$_2$ and LiFeAs single crystals: Interplay between spin fluctuations and charge transport in iron-pnictides}
\author{F. Rullier-Albenque}
\email{florence.albenque-rullier@cea.fr}
\affiliation{Service de Physique de l'Etat Condens\'e, Orme des Merisiers, CEA Saclay (CNRS URA 2464), 91191 Gif sur Yvette cedex, France}
\author{D. Colson}
\affiliation{Service de Physique de l'Etat Condens\'e, Orme des Merisiers, CEA Saclay (CNRS URA 2464), 91191 Gif sur Yvette cedex, France}
\author{A. Forget}
\affiliation{Service de Physique de l'Etat Condens\'e, Orme des Merisiers, CEA Saclay (CNRS URA 2464), 91191 Gif sur Yvette cedex, France}

\begin{abstract}
The longitudinal in-plane magnetoresistance (LMR) has been measured in different 
Ba(Fe$_{1-x}$Co$_{x}$)$_{2}$As$_{2}$ single crystals and in LiFeAs. For all these compounds, we 
find a negative LMR in the paramagnetic phase whose magnitude increases as $H^2$. We show that this negative LMR 
can be readily explained in terms of suppression of the spin fluctuations by the magnetic field. In the 
Co-doped samples, the absolute value of the LMR coefficient is found to decrease with doping 
content in the paramagnetic phase. The analysis of its $T$ dependence in an itinerant nearly antiferromagnetic 
Fermi liquid model evidences that the LMR displays a qualitative change of $T$ variation with increasing Co content. The latter occurs at optimal doping for which the antiferromagnetic ground state is suppressed. The same type of 
analysis for the negative LMR measured in LiFeAs suggests that this compound is on the verge of magnetism.

\end{abstract}

\date{\today}
\pacs{74.70.Xa, 74.25.fc, 72.15.Gd, 72.10.Di}
\maketitle

\section{Introduction}
The proximity and/or coexistence of antiferromagnetism and superconductivity (SC) in the iron-pnictides has been
taken as an indication that the magnetic fluctuations may play a decisive role in the SC pairing mechanism
\cite{Paglione,Johnston}. In the undoped compounds the spin density wave (SDW) ordering is widely attributed to 
the nesting between electron and hole quasi-cylindrical bands, which leads to a strongly peaked spin susceptibility 
at the antiferromagnetic (AF) wave vector $Q_{AF}$. It was argued very early that SC could be also mediated by 
spin fluctuations at the same wave vector, resulting in an extended s-wave pairing with sign change of 
the order parameter between electron and hole sheets \cite{Mazin2}. 

Indeed spin excitations were detected around $Q_{AF}$ by Inelastic Neutron Scattering (INS) measurements in 
the magnetic and paramagnetic (PM) states of several superconducting iron-pnictides \cite{Parshall, Lester, Diallo, Inosov}. On the other hand, nuclear magnetic resonance (NMR) measurements 
have evidenced a strong increase of 
the nuclear spin-lattice relaxation rate at low $T$ that results from the growth of AF spin fluctuations with
decreasing $T$ \cite{Ning1,Ning2}. 

Surprisingly there is not so far any identified signatures of spin fluctuations in the transport properties of iron
pnictides, which opened questions about the interplay between magnetism and charge carriers. Depending on the compounds, different power laws for the temperature dependence of the resistivity have been
observed, including linear T dependence of the resistivity for some materials. However, the multiband character of
these compounds makes it difficult to extract the $T$ dependence of the scattering rates straightforwardly.
Even if transport is dominated by one type of carriers, as in the Co substituted BaFe$_2$As$_2$, 
one has to care about a possible $T$ dependence of the number of carriers, as anticipated from the analysis of the transport properties \cite{FRA-transport} and observed recently by Angle Resolved Photoemission (ARPES) measurements \cite{Dhaka,Brouet}. Whether or not the spin fluctuations play a role in the transport properties of pnictides thus remains an interesting and open issue.

In this paper, we show that high resolution measurements of the longitudinal magnetoresistance (LMR) provide a useful probe for studying the coupling between the charge carriers and the spin degrees of freedom in prototypical iron arsenic materials. While a large positive LMR component is found in the AF phase of the pure BaFe$_2$As$_2$, a negative and quite small component is measured in the PM state of a series of Co-substituted BaFe$_2$As$_2$ and of the stoichiometric LiFeAs. The dependence of the LMR coefficient with $T$ and Co content mimics the behavior of the NMR relaxation rates both for Ba(Fe$_{1-x}$Co$_{x}$)$_{2}$As$_{2}$ and LiFeAs \cite{Ning2,Ma}. This leads us to interpret the negative LMR by the suppression of AF spin fluctuations by the magnetic field. 
We propose a qualitative analysis of the LMR in an itinerant nearly antiferromagnetic Fermi liquid model which allows us to unveil different $T$ dependences for the magnetic and non-magnetic Ba(Fe$_{1-x}$Co$_{x}$)$_{2}$As$_{2}$ samples. Our findings uncover a subtle modification of the interaction between the conduction electrons and the fluctuating magnetic moments in the paramagnetic state, which occurs simultaneously with the loss of long range antiferromagnetic order at lower temperatures when $T_c$ is optimal. Understanding these charge transport properties should give useful keys for elucidating the role of magnetic interactions in the physics of pnictides.

\section{C\lowercase{o}-doped B\lowercase{a}F\lowercase{e}$_2$A\lowercase{s}$_2$.}
\subsection{Transversal and longitudinal magnetoresistances}
The single crystals of Ba(Fe$_{1-x}$Co$_{x}$)$_{2}$As$_{2}$ were grown using the self-flux method as 
detailed in ref.\cite{FRA-transport}. The seven different Co concentrations studied, spanning the phase diagram 
from $x=0$ up to the overdoped $x \sim 0.15$, are indicated in the inset of Fig.\ref{Fig.coeff-MR}. Transport measurements were performed on samples
which have been cleaved from larger crystals to thicknesses lower than $30$
$\mu m$. Low resistivity contacts were done using silver epoxy in a four-probe configuration geometry. 
The respective values of the antiferromagnetic and superconducting transition temperatures are summarized in Table\ref{Table_samples}.

\begin{table}[h]
\caption{Characteristics of the different Ba(Fe$_{1-x}$Co$_{x}$)$_{2}$As$_{2}$ single crystals}
\label{Table_samples}
\begin{ruledtabular}
\begin{tabular}{cccccccc}
x &  0  & 0.047  &  0.06 & 0.075 & 0.1 & 0.12 & 0.154 \\
\hline
$T_N$ (K) &  137.5  &  63  &  33(1) & - & - & - & - \\
$T_c$ (K) &  - &  13.7  & 23 & 24.7 & 20.5 & 16.2 & $\sim 2.5$ \\
\end{tabular}
\end{ruledtabular}
\end{table}

The magnetoresistance (MR) measurements were done in 
fixed temperature by sweeping the magnetic field from -14T to 14T and taking the symmetric part of the signal in 
order to eliminate any spurious Hall effect component due to the misalignment of the contacts. 
As the MR is rather small, 
usually less than 0.05\% at 14T for $T>100$K, it is absolutely necessary to ensure that the temperature remains
constant during the magnetic field ramp. We have used a cernox sensor which has been calibrated in magnetic field 
and the temperature is then regulated by compensating the effect of field. The in-plane resistivity 
measurements were performed in three different geometrical configurations, with $H$ either perpendicular or parallel 
to the ab-plane. In this latter configuration we have compared the data obtained for both $H$ parallel and
perpendicular to the electrical current $J$ at selected temperatures.  

\begin{figure}
\centering
\includegraphics[width=6cm]{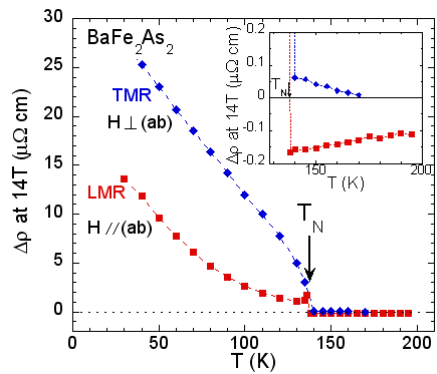}
\caption{(color on line) In-plane resistivity increase measured at 14T for BaFe$_2$As$_2$ in two different configurations (TMR:triangles, LMR:squares). A discontinuity of the two components occurs at the magnetic transition temperature $T_N$ above which both the TMR and the LMR becomes very small. Their evolution in the paramagnetic state is better visualized in the inset where the $\Delta\rho$ scale was multiplied by a factor $\sim35$.}
\label{Fig.drho-undoped}
\end{figure}

To date, MR measurements in undoped and Co-doped BaFe$_2$As$_2$ have been exclusively focused on the magnetic phase  and the signal in the paramagnetic phase has been ignored as being barely measurable (less than 0.05\% at 14T) \cite{Kuo, Ishida, Wen-high fields}. The $T$ evolution of the resistivity increase measured at 14T is reported for pure BaFe$_2$As$_2$ in Fig.\ref{Fig.drho-undoped} in the magnetic and paramagnetic phase for $\textbf{H}\perp\textbf{ab}$ and $\textbf{H}\parallel\textbf{ab}$. Both the transverse (TMR) and longitudinal (LMR) magnetoresistances undergo a discontinuity at $T_N$ and indeed the signals become very weak in the paramagnetic phase, albeit still measurable as shown in the inset of Fig.\ref{Fig.drho-undoped}. One can see there that, while the TMR remains positive and vanishes very rapidly, the LMR becomes negative and has a weaker $T$ dependence.

\begin{figure}
\centering
\includegraphics[width=8cm]{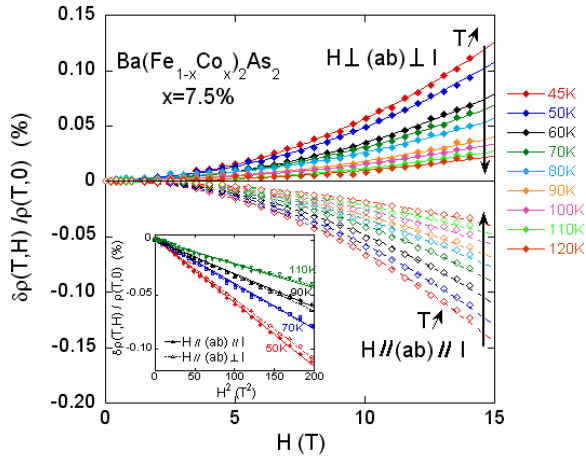}
\caption{(color on line) (a) Magnetic field dependence of the resistivity of
Ba(Fe$_{0.925}$Co$_{0.075}$)$_{2}$As$_{2}$ single crystals for temperature ranging from 30 to 120K in the 
configuration $\textbf{H} \perp (ab$-plane (top) and $\textbf{H} \parallel ab$-plane $\parallel I$(bottom). Lines are fits with a quadratic field dependence. 
Inset: The longitudinal $\delta\rho/\rho_0$ values plotted versus $H^2$ for $\textbf{H} \parallel \textbf{I}$ 
(full symbols) or $\textbf{H} \perp \textbf{I}$ (empty symbols) show that the LMR is isotropic 
in this 7.5\%Co-doped sample.}
\label{Fig.MR-H}
\end{figure}

Similar LMR and TMR behaviors are observed for all the samples in the paramagnetic phase. This is illustrated for the 
7.5\%Co-substituted sample in Fig.\ref{Fig.MR-H} that shows the field dependence of the in-plane magnetoresistance normalized to the zero 
field value $\Delta\rho/\rho_0$ for $T=45$K up to 120K. As in the PM state of the undoped compound, the TMR is positive and the LMR is negative. 
Both components show a quadratic dependence with 
the applied magnetic field with comparable absolute values which are quite small
even at low $T$ ($\mid\delta\rho/\rho_0\mid$ less than 0.2\% at 45K and 14T) and decrease with 
increasing temperatures. We can notice in the inset of Fig.\ref{Fig.MR-H} that the negative LMR does not depend 
on the respective orientations of $\textbf{H}$ and $\textbf{I}$. Very comparable LMR data are found for the underdoped  and overdoped samples 
as shown in Appendix A. As no magneto-orbital effect is expected to occur
as long as the applied magnetic field is strictly parallel to the FeAs planes in these nearly two-dimensional
compounds, this is a strong indication that the negative longitudinal components originate from a spin scattering effect. 

\subsection{Variation of the LMR with $T$ and C\lowercase{o}-doping}
The natural explanation for the negative LMR is a suppression of the spin fluctuations by the magnetic field, which results in the decrease of the magnetic part of the resistivity $\rho_{sf}(T)$. The small value of the LMR measured here suggests that this magnetic term  plays a minor role in 
the total temperature dependence of the resistivity. This is in agreement with our previous analysis of the 
transport properties of Ba(Fe$_{1-x}$Co$_{x}$)$_{2}$As$_{2}$ in which we have shown that the electron scattering 
rate displays a very similar $T^2$ dependence for all $x$ between 0.04 and 0.2, with no apparent incidence of the 
spin fluctuations \cite{FRA-transport}. 

\begin{figure}
\centering
\includegraphics[width=8cm]{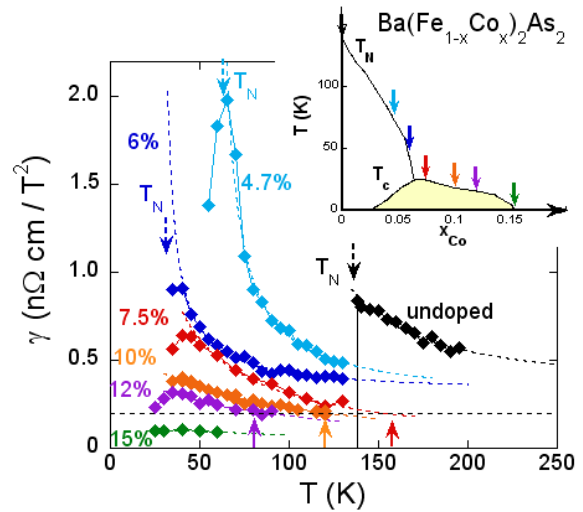}
\caption{(color on line) Temperature dependence of the longitudinal MR coefficients $\gamma(T)=-\Delta\rho/H^2$ for all the Ba(Fe$_{1-x}$Co$_x$)$_2$As$_2$ samples studied (their respective positions in the phase diagram are indicated by arrows in the inset). The shape of the curves resemble those found for the NMR $1/T_1T$ relaxation rate \cite{Ning2}, with the arrows at the bottom indicating the temperatures below which upturns of $1/T_1T$ are observed. The dashed lines are fits of the data as explained in the text.}
\label{Fig.coeff-MR}
\end{figure}

In order to compare the variation of $\rho_{sf}(T,H)$ in the different samples, we have reported in 
Fig.\ref{Fig.coeff-MR} the coefficient $\gamma(T)$ 
defined as
\begin{equation}
\Delta\rho(T,H)=-\gamma(T)H^2 
\label{eq.gamma}
\end{equation}
While $\gamma$ is barely measurable in the most doped 15.4\%Co sample with $T_c \simeq0 $, it increases 
with underdoping and reaches its maximal value for the 4.7\%Co sample near the magnetic transition. 
For all the superconducting 
samples, one can notice that $\gamma$ displays a maximum value around 40K before decreasing when approaching $T_c$. 
This might be due to the suppression of the superconducting fluctuations by the magnetic field, which would
add an additional positive contribution to the LMR below 40K. One can also see a net change of behavior between 
the 6\%Co magnetic sample and the 7.5\%Co non-magnetic sample, $\gamma$ vanishing more rapidly with increasing $T$ 
in the latter case.

The evolution of $\gamma$ with doping and temperature bears striking similarities with the behavior of the NMR
nuclear spin-lattice relaxation rate $1/T_1T$ for the same series of Ba(Fe$_{1-x}$Co$_{x}$)$_{2}$As$_{2}$ samples
\cite{Ning2}. In particular the temperatures at which upturns of $1/T_1T$ are detected in ref.\cite{Ning2}, 
indicated by full arrows at the bottom of fig.\ref{Fig.coeff-MR}, correspond very well to the onset temperature 
for observing a LMR coefficient larger than $\sim 2\ 10^{-4}\mu\Omega$cm/$T^2$. This strongly supports the relation between the negative LMR and the AF spin fluctuations. This also confirms that the superconductivity may be linked to the growth of AF spin fluctuations in these iron pnictides \cite{Ning2,Luo}  

\subsection{Analysis of the negative magnetoresistance}
In order to explain the magnetic order at $Q_{AF}$, both local-moment and itinerant scenarios have been invoked. However the observation that the local susceptibility around $Q_{AF}$ displays a Curie-Weiss behavior, while the uniform susceptibility does not, seems to favor a description in terms of itinerant electrons \cite{Johnston}. The analysis of the spin dynamics of the parent and Co-doped 122 compounds as determined by INS or NMR have shown that the imaginary part of the dynamic spin susceptibility $\chi^"(q,\omega)$ can be very well
described in the framework of a 2D itinerant nearly AF metals theory \cite{Inosov, Diallo, Ning2}. In particular, the NMR $(1/T_1T)$ has been interpreted as being directly related to the AF spin fluctuations and expressed in terms of  the static staggered susceptibility which displays the Curie-Weiss $T$ dependence $\chi_Q\propto1/(T-\theta)$, with $\theta \sim T_N$ for the magnetic samples. Negative values of $\theta$ were found for the non magnetic samples with short-range AF spin fluctuations.

The electrons that are more scattered by spin fluctuations are the ones near the hot spots of the Fermi surface, 
i.e. those connected by the ordering wave vector $Q_{AF}$. The resistivity contribution due to spin 
fluctuations $\rho_{sf}$ is thus directly expressed in terms of the dynamical spin
susceptibility $\chi^"(q,\omega)$ \cite{Moriya}. When the spin fluctuation spectrum is two-dimensional, as observed 
in the PM state of the iron pnictides \cite{Lester,Inosov,Diallo}, one finds that 
$\rho_{sf}(T) \sim T$ \cite{Moriya-cuprates, Hlubina}. The effect of a magnetic field then arises through the 
coupling between 
the uniform mode at $q=0$ and the modes with $q \approx Q_{AF}$ \cite{Usami-Moriya}. This results in a negative 
MR when this mode-mode coupling is positive. In a first approximation, the MR value can be obtained by 
considering the effect of $H$ only on the dynamical susceptibility as 
$\chi^"_H(q,\omega)=\chi^"_0(q,\omega)[1-a(T)H^2]$ where $a(T)$ measures
the mode-mode coupling effectiveness \cite{Usami}. So the magnetoresistivity due to the suppression of 
spin fluctuations can be simply written as:
\begin{equation}
\Delta\rho(T, H) \sim  - a(T) \rho_{sf}(T) H^2 \sim - a(T) T H^2
\label{Eq.MR-itinerant}
\end{equation}

\begin{figure}
\centering
\includegraphics[width=8cm]{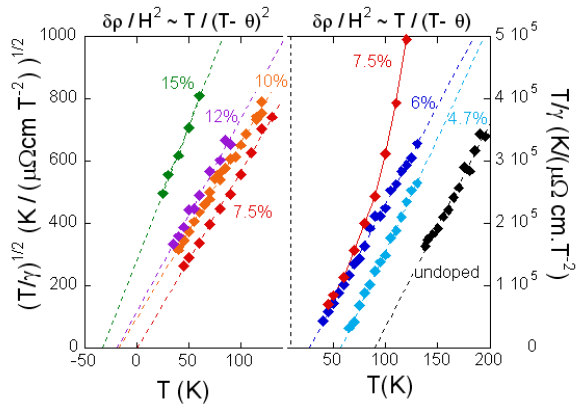}
\caption{(color on line) In the right part, the quantity $T/\gamma$ shows a linear $T$ dependence for the magnetic Co-doped 122 samples showing that the LMR can be expressed as $T/(T-\theta)$. For the non-magnetic samples (left part) the LMR is better fitted by $T/(T-\theta)^2$, as seen from the linear $T$ dependence of $\sqrt{T/\gamma}$.}
\label{Fig-fit_MR}
\end{figure}

The similarity between the behavior of $\gamma(T)$ displayed in fig.\ref{Fig.coeff-MR} and that of $1/T_1T$ suggests
that $a(T)$ is directly connected to the $T$ dependence of the staggered susceptibility. In order to test 
this assumption we have plotted in the right part of fig.\ref{Fig-fit_MR} the quantity $T/\gamma$ versus $T$ for 
the lower Co dopings (from 0 to 7.5\%). For all the magnetic samples, we observe a very good linear $T$ dependence,
which indicates that indeed $\Delta\rho(T,H)/H^2 \propto T/(T-\theta)$ with $\theta \sim T_N$ for 
$x=$ 0.047 and 0.06 ($\theta=57$ and 27K compared to $T_N= 63$ and 33K as determined from the sharp anomaly in the $\rho(T)$ curves -see Table\ref{Table_samples}). For the undoped compound, the first order of the magnetic transition prevents a direct comparison. 

However, the data for the non-magnetic 7.5\% clearly deviates from this linear behavior. 
Instead, we rather find that the MR curves can be very well scaled  by $H^2/T$ from 45 to 130K. This is visualized 
in the left part of fig.\ref{Fig-fit_MR} in which the quantity $\sqrt{T/\gamma}$ is now plotted versus $T$ for the 
non-magnetic samples. It is seen that the LMR can be written as $\Delta\rho(T,H)/H^2 \propto T/(T-\theta)^2$ where $\theta$ decreases slightly from 0 in the optimally doped sample to $\sim-50$K in the most doped 15.4\%Co one. 
These negative values of $\theta$ that reveal the existence of short-range AF spin fluctuations are in relatively 
good agreement with those extracted from INS or NMR experiments, albeit a little smaller \cite{Inosov, Ning2}.

So this analysis establishes that the negative LMR measured in the 
Ba(Fe$_{1-x}$Co$_{x}$)$_{2}$As$_{2}$ single crystals can be rather well interpreted in terms of suppression of 
the spin fluctuations by magnetic field in an itinerant electrons approach. More importantly, we also evidence
\textit{a change of the $T$ dependence of the LMR coefficient at the boundary between magnetic and non-magnetic
compounds}, which suggests that the coupling between the charge carriers and the spin fluctuations must be different 
in these two types of samples.

\section{Spin fluctuations in L\lowercase{i}F\lowercase{e}A\lowercase{s}}
\begin{figure}
\centering
\includegraphics[width=6cm]{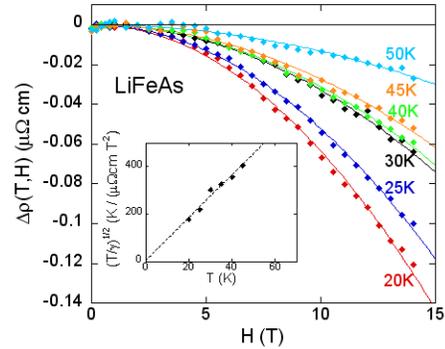}
\caption{(color on line) LMR curves for LiFeAs at different temperatures. The full lines are $H^2$ fits of the data. Inset: Plot of $\sqrt{T/\gamma}$ versus $T$. The linear extrapolation towards zero indicates that the Curie-Weiss temperature is nearly zero in LiFeAs as found for the optimally doped Co-BaFe$_2$As$_2$.}
\label{Fig-LiFeAs}
\end{figure}

The LiFeAs crystal with $T_c=16.5$K at the midpoint of the transition and a residual resistivity of 4.5$\mu\Omega.cm$ has been grown as detailed in ref.\cite{FRA-LiFeAs}. The LMR is reported for $H$ $\parallel$ $I$ in fig.\ref{Fig-LiFeAs} from 20K to 50K. Although the value of the LMR
coefficient is relatively large at low $T$ and similar to that found in optimally doped Co-BaFe$_2$As$_2$, it 
decreases much more rapidly with increasing $T$ and becomes nearly unmeasurable above 50K, as found in ref.\cite{Ma}.
Let us notice that, contrary to what is found for the Co-122, 
the spin fluctuation contribution to the MR is very small with respect to the transverse MR. Consequently it was quite
legitimate to neglect the spin contribution and assimilate the orbital MR to the TMR as done in ref.\cite{FRA-LiFeAs}. 

Even though it is not possible to discriminate between different $T$ dependences with the very few data points obtained
here, it seems natural to try to fit the data in the same way as in the optimally and overdoped Co samples \cite{footnote2}. 
This is done in the inset of fig.\ref{Fig-LiFeAs} which shows that the LMR can be expressed as 
$\Delta\rho(H,T)\propto -H^2/T$ with a Curie-Weiss temperature $\theta\sim0$, similarly to what is found for the 7.5\%Co sample. So this can be taken
as the sign that LiFeAs is located very near the boundary with the magnetic phase and that AF spin fluctuations could
play a significant role in the superconductivity of this compound. This is in agreement with the INS observation of 
a peak in the imaginary part of the susceptibility at an incommensurate AF wave vector \cite{Qureshi} despite the 
poor nesting between the electron and hole pockets observed by ARPES \cite{Borisenko}.

\section{Discussion and Conclusion}
We demonstrated here that high sensitivity measurements of the longitudinal magnetoresistance are an efficient tool
to reveal the effect of spin fluctuations on the transport properties of electron doped iron pnictides. The small and similar magnitude of the 
LMR signals found in the Co substituted BaFe$_2$As$_2$ compounds and in LiFeAs supports the idea that the effect of 
spin fluctuations cannot be directly deduced from the $T$ dependences of the resistivity, which are very different in 
these two families. Our study through the whole phase diagram of the Co substituted BaFe$_2$As$_2$ compounds allowed us to show that the coupling 
between charge carriers and spin fluctuations in the PM state is changing concomitantly with the loss of long range magnetic ordering. 

An important feature of the iron pncitides is their multiorbital nature. In particular it has been pointed out
early on that Hund's rule interaction plays a prominent role in the physics of these compounds 
by allowing a strong orbital differentiation between the $3d$ Fe orbitals \cite{Haule,Medici1,Yin2}. 
As a result, a description in terms of
coexisting itinerant and localized electrons was proposed in different theoretical models \cite{Medici1,Wu,Yin}. 
The situation appears
then reminiscent of that encountered in \textit{s-d} AF metals where the electrons in the conduction $s$ band are the 
charge carriers while those in the narrow $d$ band contribute to the spin fluctuations \cite{Moriya}. 
In this case, the spin susceptibility $\chi_c(T)$ of the conduction electrons can be considered to be Pauli-like 
and does not contribute to the $T$ dependence of the resistivity \cite{Ueda}. 

Beyond optimal doping, our experimental finding suggests
an additional $1/(T-\theta)$ contribution to the LMR upon Co doping. A possible explanation might be that 
$\chi_c(T)$ has now acquired the same $T$ dependence as the staggered susceptibility. This could be linked to 
the weakening of the orbital differentiation with electron doping predicted theoretically \cite{Yin2,Medici2}, 
which results in a strong hybridization between local moment and itinerant electrons. A salient feature is that this effect happens 
at the very boundary between magnetic and non magnetic samples where $T_c$ is optimal. This should trigger more
theoretical work, with a precise consideration of the characteristics and nesting properties of the different Fermi sheets, in order to explain the quantitative evolution of the LMR and clarify the exchange interactions between charge carriers and magnetic fluctuations in these multi-orbital systems.

As for LiFeAs, the observation of a $T$ dependence of the LMR very similar to that found in the optimally Co 
doped compound suggests that this compound is not so far from a hypothetical magnetic phase.

\section{Acknowledgments}
We thank H. Alloul for fruitful discussions and the critical reading of the manuscript. Financial support from the ANR grant "PNICTIDES" is acknowledged. 

\appendix

\section{LMR in the underdoped 4.7\%C\lowercase{o} and overdoped 10\%C\lowercase{o} substituted samples} 
\begin{figure}[h]
\centering
\includegraphics[width=7cm]{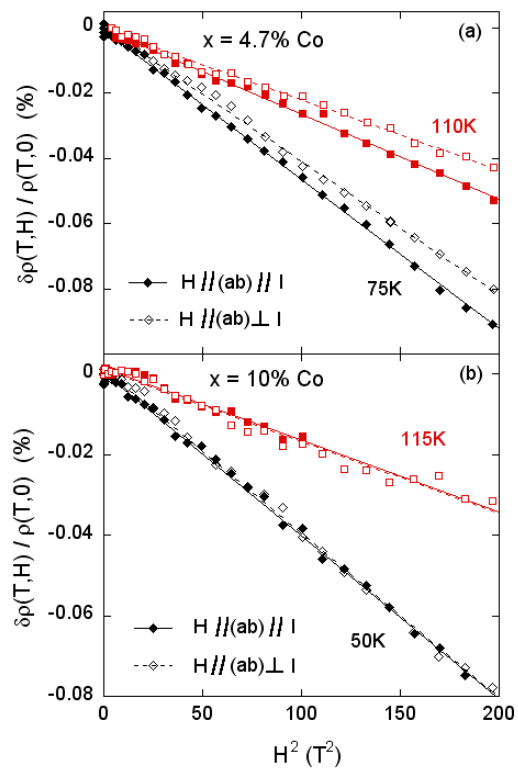}
\caption{(color in line) Longitudinal $\delta\rho/\rho_0$ values plotted versus $H^2$ for $\textbf{H} \parallel \textbf{I}$ 
(full symbols) or $\textbf{H} \perp \textbf{I}$ (empty symbols) for (a) the underdoped 4.7\%Co sample and (b) the overdoped 10\%Co sample}
\label{Fig_anisotropy}
\end{figure}

The LMR data normalized to their zero field values are plotted versus $H^2$ in Fig.\ref{Fig_anisotropy} for x=0.047 and x=0.10. As in the 7.5\%Co sample, a very good matching of the curves is observed in the overdoped sample for two respective orientations of $\textbf{H}$ and $\textbf{I}$, indicating that the LMR is isotropic in the (ab) plane. For the 4.7\%Co sample, the LMR data are found systematically larger in absolute value for $\textbf{H} \parallel \textbf{I}$ than for $\textbf{H} \perp \textbf{I}$. We do not have any clear explanation for this discrepancy, but as the ratio (of the order of 1.2) between these two configurations remains constant with temperature, this will not result in different temperature dependences of the LMR components. So we have only considered here the LMR component measured for $\textbf{H} \parallel \textbf{I}$ for which only the spin effect is expected to contribute to the magnetoresistivity.

\begin{thebibliography}{99}
\bibitem{footnote2}Fitting the data as $T/\gamma$ versus $T$ as done for the underdoped Co-doped samples would indeed also result in a rather good linear $T$ dependence but with a value of $\theta$ equal to 15K, which appears unphysical for this non-magnetic sample.
\end{thebibliography}
\end{document}